\documentclass[twocolumn]{autart}
\usepackage{epsfig}
\usepackage{epstopdf}
\usepackage{graphicx}
\usepackage[mathscr]{eucal}
\usepackage{amsmath,amsfonts,amssymb,amsbsy,amsopn,amstext}
\usepackage{latexsym}
\usepackage{color,multicol,makeidx}
\usepackage{graphicx}
\usepackage{bm}
\usepackage{mathrsfs}
\usepackage{float}
\usepackage{ccaption}
\usepackage{cite}
\usepackage{natbib}
\usepackage{mathdots}
\usepackage{algorithm}
\usepackage{algpseudocode}
\usepackage{pifont}
\usepackage[caption=false,font=normalsize,labelfont=sf,textfont=sf]{subfig}
\usepackage{enumerate}
\usepackage{tabu}
\usepackage{listings}
\def\dref#1{(\ref{#1})}
\def\eq{\displaystyle\stackrel\triangle=}

\newcommand{\diag}{\ensuremath{\mathrm{diag}}}
\DeclareMathOperator*{\rank}{rank} 
\DeclareMathOperator*{\argmin}{arg\,min}

\begin{document}
	
	\begin{frontmatter}
		
		\title{On Embeddings and Inverse Embeddings of Input Design for Regularized System Identification}
		\vspace{-3mm}
		\author[AMSS]{Biqiang Mu}\ead{bqmu@amss.ac.cn},
				\author[CUHK]{Tianshi Chen}\ead{tschen@cuhk.edu.cn},
		\author[SUST1,SUST2]{He Kong}\ead{kongh@sustech.edu.cn},
		\author[NNU]{Bo Jiang}\ead{ jiangbo@njnu.edu.cn},
		\author[ZJU]{Lei Wang}\ead{lei.wangzju@zju.edu.cn},
		\author[CUHK,ZJU]{Junfeng Wu}\ead{junfengwu@cuhk.edu.cn}

		\address[AMSS]{Key Laboratory of Systems and Control, Institute of Systems Science,
			Academy of Mathematics and Systems Science, Chinese Academy of Sciences, Beijing~100190, China}
		\address[CUHK]{School of Data Science and Shenzhen Research Institute of Big Data, The Chinese University of Hong Kong, Shenzhen 518172, China}

		\address[SUST1]{Shenzhen Key Laboratory of Biomimetic Robotics and Intelligent Systems, Department of Mechanical and Energy Engineering, Southern University of Science and Technology, Shenzhen, 518055, China}
				\address[SUST2]{Guangdong Provincial Key Laboratory of Human-Augmentation and Rehabilitation Robotics in Universities, Southern University of Science and Technology, Shenzhen, 518055, China.}
		
				\address[NNU]
		{Key Laboratory for NSLSCS of Jiangsu Province, School of Mathematical Sciences, Nanjing Normal University, Nanjing, China}
				\address[ZJU]{College of Control Science and Engineering, Zhejiang University, Hangzhou, China}
		
		\begin{abstract}
			Input design is an important problem for system identification and has been well studied for the classical system identification, i.e., the maximum likelihood/prediction error method. For the emerging regularized system identification, the study on input design has just started, and it is often formulated as a non-convex optimization problem that minimizes a scalar measure of the Bayesian mean squared error matrix subject to certain constraints, and the state-of-art method is the so-called quadratic mapping and inverse embedding  (QMIE) method, where a time domain inverse embedding (TDIE) is proposed to find the inverse of the quadratic mapping. In this paper, we report some new results on the embeddings/inverse embeddings of the QMIE method. Firstly, we present a general result on the frequency domain inverse embedding (FDIE) that is to find the inverse of the quadratic mapping described by the discrete-time Fourier transform. Then we show the relation between the TDIE and the FDIE from a graph signal processing perspective.
			Finally, motivated by this perspective, we further propose a graph induced embedding and its inverse, which include the previously introduced embeddings as special cases.   
			This deepens the understanding of input design from a new viewpoint beyond the real domain and the frequency domain viewpoints. 
			
		\end{abstract}
		
	\end{frontmatter}

	\section{Introduction}
	
	Over the past decade, kernel-based regularization methods (KRMs)
	have  received increasing attention in the system identification
	community \citep{Pillonetto2010,Chen2012,Pillonetto2014,Chiuso2016}, which has shown better average accuracy and robustness compared to  the
	classical maximum likelihood/prediction error methods for short/low signal-to-noise ratio data.
	The key idea of the KRM is to first encode prior knowledge on
	the dynamic system by parameterizing the kernel matrix with a few number of parameters (kernel design), called hyperparameter,  then to estimate the hyperparameter based on the data (hyperparameter estimation), and finally to calculate the regularized least squares estimator of the model.
	For kernel design, 
	many kernels have been proposed for various kinds of prior knowledge such as exponential decaying, smoothness, high-frequency decay property, direct current gain, and among others \citep{Chen2016,Chen2018a,Carli2017,Marconato2016,Zorzi2018,Pillonetto2016,Fujimoto2018c,Fujimoto2021}. 
	For hyperparameter estimation, the asymptotic properties of the empirical Bayes (EB) estimator,  the Stein's unbiased risk estimator (SURE), cross-validation (CV) estimator have been investigated in the sense of  mean square error (MSE) 
	\citep{Pillonetto2015,Mu2018a,Mu2021}.
	In particular, it was shown in
	\cite{Mu2018a,Mu2021} that  the SURE and the CV estimators are{}
	asymptotically optimal but the widely used EB estimator
	is not in the MSE  sense.
	
	Input design is an important problem in system identification and can be used to further improve the performance of model estimators by careful design of the input signal. For ML/PEM, input design has been well studied, e.g.,  the survey papers
	\citep{Mehra1974,Hjalmarsson2005,Gevers2005} and the monographs
	\citep{Goodwin1977,Ljung1999,Zarrop1977}).	
	The classical method of input design, e.g., \cite{Jansson2005,Hildebrand2003,Hjalmarsson2009}, is to minimize a scalar measure (e.g., the determinant, the trace or others) of the \emph{asymptotic} covariance matrix of the parameter estimators with constraints on the input and/or output, which can be considered both in the time domain and in the frequency domain. 
	In contrast to the input design in the time domain, the input design in the frequency domain has a clear physical interpretation \citep{Jansson2005}.
	
	
	For KRM, the study on input design has just started \citep{Fujimoto2018,Mu2018b,Fujimoto2018b}.
	This issue was first investigated for a fixed kernel matrix in \cite{Fujimoto2016,Fujimoto2018} by maximizing the mutual information between the output and the impulse response subject to an input energy constraint.
	Since the formulated optimization problem in \cite{Fujimoto2018} is non-convex, projected gradient algorithms are adopted to search for optimal inputs.
	This issue was also considered in the frequency domain \citep{Fujimoto2018b} by minimizing the posterior uncertainty of the model with a bounded input constraint, where the input sequence was generated by an online greedy algorithm  and also an offline algorithm  based on a projected gradient method was employed to search for the optimal input.
	
	Different from  \cite{Fujimoto2018,Fujimoto2018b},
	the input design problem considered in \cite{Mu2018b} was formulated as a non-convex optimization problem that minimizes a scalar measure (the determinant, the trace, or the largest eigenvalue) of the Bayesian mean squared error (the posterior covariance of the parameter estimate) subject to the input power constraint. Moreover, a quadratic mapping and inverse embedding  (QMIE) method was proposed to search for optimal inputs and in particular, 
	the quadratic mapping (that is expressed by a composition of three simple real mappings, called the \emph{time domain embedding} (TDE) since it works in the time domain) from the input to its autocovariance transforms the non-convex optimization problem (with respect to the input) to a convex one (with respect to the autocovariance), and the inverse image set of the quadratic mapping from  given autocovariance to its associated inputs is explicitly characterized and called the \emph{time domain inverse embedding} (TDIE).  That is, the QMIE method first calculates the optimal autocovariance and then finds optimal inputs by invoking the  TDIE. 

	Interestingly, as pointed out by an anonymous reviewer of \cite{Mu2018b}, the idea to use embeddings and inverse embeddings for input design problems has actually appeared before in classical system identification. In particular, the \emph{frequency domain embedding} (FDE) of the quadratic mapping described by the discrete-time Fourier transform (DFT) and the corresponding inverse embedding (describing the inverse image set of the quadratic mapping for a given autocovariance by the FDE), called the \emph{frequency domain inverse embedding} (FDIE), have been sketched in \cite{Hjalmarsson2006,Jansson2004}. 
	However, it seems that  the proposed FDIE only works for the case $N=n$, where $N$ and $n$ are the sample size and the number of model parameters, respectively, but can not be directly applied to the more general case $N>n$.	Moreover, the FDIE only sketches a route to find inputs for given squared magnitudes of the DFT of the input, but does not give a complete description of inputs for a given autocovariance, i.e., the inverse image set of the quadratic mapping).
	Then it is natural to ask the following problems:
	\begin{itemize}
		
				\item Is it possible to give a complete description of the inverse image set of the quadratic mapping for given autocovariance based on the FDE and especially for the case $N>n$?

		\item What is the relation between the FDIE in \citep{Hjalmarsson2006,Jansson2004} and the TDIE in \cite{Mu2018b}?
	\end{itemize}

	
	

	In this paper, we aim to address these problems. In particular, we first study how to characterize the FDIE and then to establish the relation between the TDIE and the FDIE and we show that for the given autocovariance, both the TDIE and the FDIE give the same set of inputs. Interestingly, this finding can actually be interpreted from a unified graph signal processing perspective, e.g., \Citet{Sandryhaila2013}, and moreover, motivated by this perspective, we further propose a graph induced embedding and its inverse, which include the previously introduced TDE and FDE as special cases. 
	The graph signal processing perspective provides a new viewpoint on understanding the input design problem. Also,  some well  developed  tools for graph signal processing \citep{Sandryhaila2013}, e.g., graph Fourier transform, graph spectral representation, etc, might have the potential to be used for input design problems.
	Finally, it is worth to note that the obtained results also applies to the case without regularization, i.e.,  the least squares estimators.



	The remaining parts of this paper are organized as follows. In
	Section \ref{sec2}, we first briefly review  the input design problem of the KRM and then present the problem statement. In Section
	\ref{sec3},  we present an explicit route to fully characterize the FDIE for the case $N\geq n$.
	In Section \ref{secgraph}, we first study the relation between TDE and FDE and then interpret them from a unified graph signal processing perspective, which motivates us to find more embeddings. Finally, we conclude the paper in Section \ref{con}. All proofs of
	Theorems and Propositions are postponed to the Appendix.

	\section{Preliminaries and Problem Statement}
	\label{sec2}
	\subsection{Regularized Least Squares Estimators}
	Consider a discrete-time time-invariant finite impulse response (FIR)  system
	\begin{align}
	y_t = a_1 u_{t-1} +\cdots+a_nu_{t-n} +  \epsilon_t,~1\leq t \leq N,\label{sys}
	\end{align}
	where $ y_t,u_t\in\mathbb R$ are the
	output and input of the system at time $t$, respectively,
	$\{\epsilon_t\}$ is a sequence of zero mean white noise with finite variance
	$\sigma^2>0$ and is independent of input $\{u_t\}$.
	The system (\ref{sys}) has the following matrix-vector form:
	\begingroup
	\allowdisplaybreaks
	\begin{subequations}
		\begin{align}
		y&=\Phi \theta + \epsilon, ~\theta=[a_1~a_2~\cdots~a_n]^T,\label{fir3}\\
		\Phi&=
		\begin{bmatrix}
		u_0&u_{-1}&\cdots&u_{-n+1}\\
		u_{1}&u_{0}&\cdots&u_{-n+2}\\
		\vdots&\vdots&\vdots&\vdots\\
		u_{N-1}&u_{N-2}&\cdots&u_{N-n}\\
		\end{bmatrix},
		\\
		y&=
		[y_{1}~y_2~\cdots~y_N]^T,
		\epsilon=[\epsilon_{1}~\epsilon_2~\cdots~\epsilon_N]^T,
		\label{eq:regressM}
		\end{align}
	\end{subequations}
	\endgroup
	where $(\cdot)^T$ denote the
	transpose of a matrix or vector. 
	
	The least squares (LS) estimator 
	$
	\widehat{\theta}_N^{\rm LS}=\argmin_{\theta \in \mathbb{R}^n}\|y-\Phi\theta\|^2 
	= (\Phi^T\Phi )^{-1}\Phi^T y
	$
	is  a prevalent way to identify the parameter vector $\theta$.
	When either $n$ is  large or the input is ill-conditioned, the LS estimator might have a large variance \citep{Pillonetto2010,Chen2012}.
	While the regularized least squares (RLS) estimator, see e.g., \cite{Chen2012}, defined by
	\begin{subequations}\label{eq:rls}
		\begin{align}
		\widehat{\theta}_N^{\rm R}=&\argmin_{\theta \in \mathbb{R}^n}\|y-\Phi\theta\|^2 + \sigma^2\theta^TK^{-1}\theta\\
		=&(\Phi^T\Phi   + \sigma^2 K^{-1})^{-1}\Phi^Ty \label{rls}
		\end{align}
	\end{subequations}
	can mitigate the large variance problem of the LS estimator by introducing a small bias,
	where $K\in{\mathbb R}^{n\times n}$ is called a kernel matrix and assumed to be positive definite
	($\sigma^2K^{-1}$ is often called the regularization matrix).
	Also, the RLS estimator \dref{eq:rls} can be explained as the posterior mean of the parameters $\theta$ for the Gaussian prior $\theta \sim \mathscr{N}(0,K)$ in a Bayesian perspective. 
	
	Given the data $\{u_t,y_t,1\leq t\leq N\}$, to make the RLS estimator \dref{eq:rls} achieve a good performance, it is necessary and critical to tune the kernel matrix $K$ in terms of data. Kernel-based regularization methods proposed in \cite{Pillonetto2010}   established a two-step procedure to select a ``good"  $K$ by embedding prior knowledge of the system to be identified, consisting of kernel design and hyperparameter estimation.
	
	Kernel design is to parameterize the matrix $K$ by a few number of parameters $\eta$, called hyperparameter, namely,
	$
	K(\eta), ~\eta \in \Omega\subset \mathbb R^p,
	$
	and meanwhile prior knowledge of the system to be identified (exponential stability and smoothness) is encoded within the structure of $K$.
	Several parameterization strategies of $K$ have been proposed, 
	such as the stable spline (SS) kernel
	\citep{Pillonetto2010}, the diagonal correlated (DC) kernel and the
	tuned-correlated (TC) kernel \citep{Chen2012}, etc.
	
	Hyperparameter estimation is to estimate the hyperparameter $\eta$ for a given parameterization of $K$ by the data in terms of optimization criteria.
	Currently prevalent hyperparameter estimators include the empirical Bayes (EB) estimator,
	the Stein's unbiased risk estimator (SURE), cross-validation (CV) estimator, and so on \citep{Pillonetto2014}. 



	\subsection{Input-design Problem for Regularized Linear System Identification}
	
	There has been a lot of work dedicated to the classical input design issue of the FIR model \dref{sys} estimation   from both time domain and frequency domain, e.g.,  \cite{Goodwin1977,Jansson2004}. 
	While the goal of input design for the regularized FIR model estimation is to determine
	an input sequence 
	$$
	\{u_{-n+1},\cdots,u_{-1},u_0, u_1,\cdots,u_{N-1}\}
	$$
	such that the RLS estimator \dref{eq:rls} is
	as good as possible under given constraints.
	
	Since the input sequence has finite length in practice, there are mainly two ways to design inputs.
	One way is to consider  asymptotic approximation, namely minimizing a scalar measure of
	the asymptotic covariance matrix of the parameter estimate (the asymptotic covariance matrix of the estimated transfer function in frequency domain) 
	\citep{Ljung1999}. As mentioned in \cite{Jansson2004}, the asymptotic approximation might not be accurate in some cases.
	The other way is to assume the unknown initial inputs as
	\begin{equation}
	\begin{aligned}\label{eq:assoninicond}
	u_{-i} &= u_{N-i},\quad i=1,\cdots,n-1,
	\end{aligned}\end{equation}
	which was introduced in \citet{Hjalmarsson2006,Jansson2004},
	such that the input sequence 
	$
	\{u_{-n+1},\cdots,$
	$u_{-1},u_0, u_1,\cdots,u_{N-1}\}
	$
	is $N$-periodic and $\Phi$ is circulant.
	This assumption \dref{eq:assoninicond} guarantees that  the explicit expression for the covariance matrix of the parameter estimate (the estimated transfer function in frequency domain) is accurate for finite sample sizes.
	
	This paper adopts the latter way, i.e., the assumption \dref{eq:assoninicond}, and the input design problem for the RLS estimator \dref{eq:rls}  is formulated as follows:  given a tuned kernel matrix $K$ and known $\sigma^2$,
	the optimal input $u^*$ is optimized by 
	\begingroup
	\allowdisplaybreaks
	\begin{subequations}\label{op}
		\begin{align}
		u^*  &\eq \arg\min_{u\in \mathscr{U}} J (\sigma^2 P^{-1}), ~~P= \Phi^T\Phi + \sigma^2 K^{-1},\\
		\mathscr{U}&=\left\{u=[u_0\cdots,u_{N-1}]^T\in\mathbb{R}^{N}\Big|~u^Tu= \mathcal{C}\right\},  \label{cons}
		\end{align}
	\end{subequations}
	\endgroup
	where $\mathcal{C}$ is a predetermined constant (the power constraint) and 
	the function $J(\cdot)$ is concave and strictly increasing with respect to the convex cone $S_n^+$ (consisting of symmetric positive definite matrices of size $n\times n$). Namely, for $ X,Z\in S_n^+$, there should hold that
	$$\alpha J(X) + (1-\alpha) J(Z) \leq J(\alpha X + (1-\alpha) Z)$$
	for all $0\leq \alpha \leq 1$, and  $J(X)\geq J(Z)$ if $X-Z\in S_n^+$.
	When $\sigma^2 K^{-1}=0$, the problem \dref{op} reduces to the input design problem for the LS estimator (See (6.3.11)--(6.3.12) of \cite{Goodwin1977}).
	
	\begin{rem}
		The posterior covariance  of the RLS estimator \dref{eq:rls} is $\sigma^2 P^{-1}$ if the $\theta$ has a Gaussian prior $\theta \sim \mathscr{N}(0,K)$ \citep{Chen2012}.
		The concave function $J(\cdot)$ is a scalar measure of the posterior covariance $\sigma^2 P^{-1}$ and some typical choices of $J(\cdot)$ are the logarithm of determinant,  the trace,  and the least eigenvalue of a positive definite matrix, 
		which correspond to the classic $D$-optimality, $A$-optimality, and $E$-optimality, respectively \citep{Ljung1999}.	
	\end{rem}

	\subsection{Quadratic Mapping and Inverse Embedding Methods}
	
	The quadratic mapping and inverse embedding (QMIE) method proposed in \cite{Mu2018b} introduces a two-step procedure for finding global minima of the nonconvex input design problem \dref{op}. 
	Under the periodic assumption \dref{eq:assoninicond} on the unknown initial inputs, the QMIE method essentially relies on the following vertor-valued quadratic mapping $f$ from the input $u$ to its autocovariance sequence $r$, defined by
	\begin{align}
	\label{eq:defmap}r=f(u)=[f_0(u),\cdots,f_{n-1}(u)]^T
	\end{align} with $r=[r_0,r_1,\cdots,r_{n-1}]^T$ and
	\begin{align}
	r_i = f_i(u)=\sum_{t=0}^{N-1}u_tu_{t-i},~~0\leq i\leq n-1,	\label{autocor}
	\end{align} 
	where $r_0=\mathcal{C}$ since the total power constraint $\sum_{t=1}^N u_{t}^2= \mathcal{C}$.
	Thus,
	the Gram matrix 
	\begin{align}
	\Phi^T\Phi =\left[
	\begin{array}{cccccc}
	r_0&r_{1}&\cdots &r_{n-2} &r_{n-1}\\
	r_1&r_0&\ddots &r_{n-3}&r_{n-2}\\
	\vdots&\ddots&\ddots&\ddots&\vdots\\
	r_{n-2}&r_{n-3}&\ddots&r_0&r_{1}\\
	r_{n-1}&r_{n-2}&\cdots&r_1&r_0\\
	\end{array}
	\right]\label{phiphi}
	\end{align}
	is Toeplitz and positive semidefinite.
	Therefore, the original nonconvex input design problem \dref{op}
	is transformed into
	a convex problem with respective to $r$:
	\begin{align}
	r^* =\argmin_{r\in \mathscr{F}} J(\sigma^2 P^{-1}),\label{or}
	\end{align}
	where the constraint set $\mathscr{F}=\{f(u)|u\in\mathscr{U}\}$ is a convex polytope described by a group of known vertices.
	
	As a result, the first step of the QMIE method is to find a global minimum of the convex problem (\ref{or}) by convex optimization algorithms, and the second step is to find a $u\in \mathscr{U}$ for any given $r\in \mathscr{F}$, e.g., characterizing the inverse mapping $f^{-1}(\cdot)$ of the $f(\cdot)$, namely, given any $r\in \mathscr{F}$,  find the set $f^{-1}(r) \eq \{u\in\mathscr{U} | f(u)=r\}$. 
	Let 
		\begin{align}
	&\xi_j\eq \left[
	\begin{array}{c}
	1\\
	\cos(j\varpi)\\
	\vdots\\
	\cos((N\!\!-\!1)j\varpi)
	\end{array}
	\right],~ \zeta_j\eq \left[
	\begin{array}{c}
	0\\
	\sin(j\varpi)\\
	\vdots\\
	\sin((N\!\!-\!1)j\varpi)
	\end{array}
	\right]\label{xi}
	\end{align}
	with $\varpi=2\pi/N$ 	for $j\geq 0$.
	Define the matrices
		\begin{subequations}
				{\scriptsize
		\begin{align}
		&S=\Big[\xi_0,\xi_1,\cdots,\xi_{n-1}\Big]^T\in\mathbb{R}^{n\times N},\label{S}\\
		&W\!=\!\left\{\begin{array}{l}
		\sqrt{\frac{2}{N}} \left[
		\frac{\xi_0}{\sqrt{2}},\xi_1,\cdots,\xi_{\frac{N-2}{2}},\frac{\xi_{\frac{N}{2}}}{\sqrt{2}},\zeta_{\frac{N-2}{2}},\cdots,\zeta_1
		\right]~\mbox{for odd $N$}\\
		\sqrt{\frac{2}{N}} \left[
		\frac{\xi_0}{\sqrt{2}},\xi_1,\cdots,\xi_{\frac{N-1}{2}},\zeta_{\frac{N-1}{2}},\cdots,\zeta_1
		\right]~\mbox{for even $N$}.
		\end{array}\right.
		\end{align}}
		\end{subequations}The characterization of its inverse mapping and inverse set is achieved in \cite{Mu2018b} by rewriting $f(\cdot)$ as a composition of three simple mappings: $f(u) = h_1(h_2(h_3(u)))$ with
	\begingroup
	\allowdisplaybreaks
	\begin{subequations}
		\label{tdd}
		\begin{align}
		h_1(z^2)&=Sz^2,  \label{md2}\\
		h_2(z)&=[z_0^2,z_1^2,\cdots,z_{N-1}^2]^T, \label{md3}\\
		h_3(u) &= W^T u, \label{md4}
		\end{align}
	\end{subequations}
	\endgroup
	where $z=[z_0,z_1,\cdots,z_{N-1}]^T$, $z^2=[z_0^2,z_1^2,\cdots,z_{N-1}^2]^T$, 
	and $W$ is an orthogonal matrix  of size $N\times N$.
	The route from $r$ to $u$ with $r\in \mathscr{F}$ based on the embedding \dref{tdd}, i.e., finding the $f^{-1}(r)$,  refers to  (53)--(55) of \cite{Mu2018b}, termed the \textit{time domain inverse embedding} (TDIE) in the following since it works in the time domain.
	Accordingly, the expression \dref{tdd} is called the \textit{time domain embedding} (TDE).
	
	For a periodic input \dref{eq:assoninicond}, it has been found in \cite{Hjalmarsson2006,Jansson2004}, the autocovariance $r$ defined in \dref{eq:defmap} can be put into the form of
	\begin{align}
	r_i 
	= \sum_{k=0}^{N-1}|U_k|^2  e^{j\varpi k i },	\label{autocorfft}
	\end{align}
	where $0\leq i\leq n-1$, $\varpi=2\pi/N$, $j$ is the imaginary unit $j^2=-1$, and
	\begin{subequations}
		\label{fftp}
		\begin{align}
		U_k&=\frac{1}{\sqrt{N}}\sum_{t=0}^{N-1}u_te^{-j\varpi kt},~k=0\cdots,N-1,	\label{fft}\\	
		u_t&=\frac{1}{\sqrt{N}}\sum_{k=0}^{N-1}U_ke^{j\varpi kt},~t=0,\cdots,N-1,
		\label{ifft}
		\end{align}
	\end{subequations}
	which are the discrete Fourier transform (DFT) pair of an $N$-periodic signal $u$. It is clear that the DFT coefficients satisfy
	$\overline{U_k}=U_{N-k},~k=1,\cdots,N-1 $, where $\overline{(\cdot)}$ means the conjugate of a complex number, vector, or matrix. Actually, the equation \eqref{autocorfft} also defines the mapping from the DFT of the input signal $u$ to $r$ and thus is also an embedding but stated in the frequency domain, and thus called the \textit{frequency domain embedding} (FDE) in the sequel. The FDE \dref{autocorfft} clearly shows that 
	the magnitude of the $k$-th spectral line of the autocovariance coefficient $r_i$ for $0\leq i \leq n-1$  at  frequency $\varpi ki$  is the squared magnitude $|U_k|^2$. 

	Moreover, it has been suggested by \cite{Hjalmarsson2006,Jansson2004}
	and also by an anonymous reviewer of \cite{Mu2018b} that the \textit{frequency domain inverse embedding} (FDIE) based on the FDE \dref{autocorfft} (finding $f^{-1}(r)$ for a given $r\in\mathscr{F}$) can be  obtained according to the following procedure:
	\begin{enumerate}[i).]
		\item take the inverse Fourier transform of $r$ and get:
\begin{align}
|U|^2\eq[|U_0|^2,\cdots,|U_{N-1}|^2]^T, \label{ifft2}
\end{align}	
\item get \begin{align}
U\eq[U_0,\cdots,U_{N-1}]^T
\end{align} by making the square-root of $|U_k|^2$, $k=1,\cdots,N-1$, and assigning phases consistent with a Fourier transform.
		\item take the inverse Fourier transform \dref{ifft} and obtain the $u$ satisfying $f(u)=r$.
	\end{enumerate}
It should be noted that the  first step \dref{ifft2} of the FDIE  involves the inverse Fourier transform. When $N>n$, we can not directly take the inverse Fourier transform of $r$ with dimension $n$ and obtain the vector $|U|^2$ with dimension $N$. Actually, it needs more technical treatments.
On the other hand, it is clear to see that the inverse embedding (finding the inverse mapping of the quadratic mapping \dref{eq:defmap}) is an essential step not only for the input design problem \dref{op} of the RLS estimator but also for the input design problem (See (6.3.11)--(6.3.12) of \cite{Goodwin1977}) of the LS estimator.  
	
	\subsection{Problem Statement}

	In this paper, we aim to investigate the embedding and inverse embedding problems for the input design problem \dref{op} and in particular, we are interested in the following questions:
	\begin{enumerate}[Q1:]
		\item How to characterize the FDIE of the mapping \dref{eq:defmap} for the more general case $N\geq n$?
		
		
		\item What is the relation between the TDE and the  FDE of the mapping \dref{eq:defmap} as well as their inverse embeddings?
		
		\item Whether or not there exist more embeddings besides \dref{tdd} and \dref{autocorfft}?
		
	\end{enumerate}
	
	It should be noted that	solutions to these questions will deepen our understanding on the input design problem \dref{op} not only for the RLS estimator and but also for  the LS estimator.

	\section{Frequency Domain Inverse Embedding}
	\label{sec3}
	In the following,  when $N\geq n$, we give an explicit way to fully characterize the FDIE in terms of the following composite decomposition of the FDE \dref{autocorfft}--\dref{fftp}:
	\begin{subequations}
		\label{tdf}
		\begin{align}
		f(u) &= \widetilde{h}_1(\widetilde{h}_2(\widetilde{h}_3(u)))~~\mbox{with} \label{mf1}\\
		\widetilde{h}_1(|U|^2)&=\widetilde{S}|U|^2, \label{mf2}\\
		\widetilde{h}_2(U)&=[|U_0|^2,\cdots,|U_{N-1}|^2]^T, \label{mf3}\\
		\widetilde{h}_3(u) &=U= \widetilde{W} u, \label{mf4}
		\end{align}
	\end{subequations}
	where 
	\begingroup
	\allowdisplaybreaks
	\begin{subequations}
		\label{sw}
		\begin{align}
		\label{tildes}
		\widetilde{S}&\eq \left[
		\begin{array}{ccccc}
		1&1&\cdots&1\\
		1&e^{j\varpi}&\cdots&e^{j\varpi(N-1)}\\
		\vdots&\vdots&\ddots&\vdots\\
		1&e^{j\varpi(n-1)}&\cdots&e^{j\varpi(n-1)(N-1)}
		\end{array}
		\right],\\
		\label{tildew}
		\widetilde{W}&\eq\frac1{\sqrt{N}}\left[
		\begin{array}{ccccc}
		1&1&\cdots&1\\
		1&e^{-j\varpi}&\cdots&e^{-j\varpi(N-1)}\\
		\vdots&\vdots&\ddots&\vdots\\
		1&e^{-j\varpi(N-1)}&\cdots&e^{-j\varpi(N-1)^2}
		\end{array}
		\right].
		\end{align}
	\end{subequations}
	\endgroup
	Moreover, the image  of $\widetilde{h}_3(\cdot)$  under $\mathscr{U}$ is
	\begin{align}
	\nonumber
	\widetilde{\mathscr{Z}}&=\{\widetilde{h}_3(u)|u\in \mathscr{U}\}\\
	&=\{U|U^HU= \mathcal{C}, \overline{U_k}=U_{N-k},~k=1,\cdots,N-1 \}, \label{image1}
	\end{align}
	where $(\cdot)^H$ denotes the complex conjugate transpose of  a complex vector or matrix,
	the image  of $\widetilde{h}_2(\cdot)$  under $\widetilde{\mathscr{Z}}$
	\begingroup
	\allowdisplaybreaks
	\begin{align}
	\nonumber
	\widetilde{\mathscr{X}}&=\{\widetilde{h}_2(U)|U\in \widetilde{\mathscr{Z}}\}\\
	\nonumber
	&
		=\Big\{|U|^2\Big|\sum_{k=0}^{N-1}|U|^2_k\! =\! \mathcal{C},|U|^2_k \geq 0,k=\!0,1,\dots,N\!-\!1,\\
	&	\hspace{15.5mm} |U_k|^2=|U_{N-k}|^2,~k=1,\cdots,N\!-\!1 \Big\}
\label{image2}
	\end{align}
	\endgroup
	is  convex,
	and the image of $\widetilde{h}_1(\cdot)$  under $\widetilde{\mathscr{X}}$
	(also the image of $f(\cdot)$ under $\mathscr{U}$)
	\begin{align}
	\mathscr{F}=\{f(u)|u\in\mathscr{U}\} =\{\widetilde{S}|U|^2 \big| |U|^2\in \widetilde{\mathscr{X}}\} \label{image3}
	\end{align}
	is a convex polytope.
	
	Based on the decomposition \dref{tdf}, the FDIE (finding the set $f^{-1}(r)$ with $r\in \mathscr{F}$) can be obtained according to the following procedure:
	\begin{enumerate}[i).]
		\item finding the inverse image of $\widetilde{h}_1(\cdot)$ for $r\in \mathscr{F}$:
		\begin{align}
		\nonumber
		\hspace{-6mm}
		\widetilde{\mathscr{X}}(r)
		\eq&\left\{|U|^2 \Big|\widetilde{S}|U|^2 =r, |U|^2 \in\widetilde{\mathscr{X} }\right\}\\
		\nonumber
		=&\Big\{ |U|^2\Big |
		\widetilde{S}|U|^2 = r,~|U_k|^2\geq 0,0\leq k \leq N-1,\\
		&~~|U_k|^2=|U_{N-k}|^2, 1\leq k \leq N \!-\!1\Big\},
		\label{b1}
		\end{align}
		where the constraint $\sum_{k=0}^{N-1}|U|^2_k\! =\! \mathcal{C}$ in $\widetilde{\mathscr{X}}$ is included by the first equality of $\widetilde{S}|U|^2 = r$;
		\item finding the inverse image of $\widetilde{h}_2(\cdot)$ for $|U|^2\in\widetilde{\mathscr{X}}(r)$:
		\begin{align}
		&\widetilde{\mathscr{Z}}(r)
		\eq\left\{ U |h_2( U) \in\widetilde{\mathscr{X}}(r)\right\}\\
		=&\left\{ \scriptsize
		\begin{array}{ll}
		\Big\{U\Big|
		U_0=\pm\sqrt{|U_0|^2},~U_{N/2}=\pm\sqrt{|U_{N//2}|^2},\\
		\nonumber
		\hspace{6mm} U_k=\sqrt{|U_k|^2}e^{j\beta_k}, ~U_{N-k}=\overline{U_k},\\
		\hspace{6mm} 0\leq \beta_k <2\pi,1\leq k \leq N/2-1
		\Big\}
		&  
		\mbox{for even } N, \\
		\Big\{U\Big|
		U_0=\pm\sqrt{|U_0|^2},U_{N-k}=\overline{U_k} \\
		\nonumber
		\hspace{6mm}U_k=\sqrt{|U_k|^2}e^{j\beta_k},
		~,0\leq \beta_k <2\pi\\
		\hspace{6mm} 1\leq k \leq (N-1)/2
		\Big\} & \mbox{for odd } N,
		\end{array}
		\right.
		\end{align}
		   where each phase $\beta_k$ can be arbitrary between $0$ and $2\pi$;
		
		\item finding the inverse image of $\widetilde{h}_3(\cdot)$ for $U\in \widetilde{\mathscr{Z}}(r)$:
		\begin{align}
		\hspace{-4mm}\widetilde{\mathscr{U}}(r)
		\eq\left\{u| \widetilde{W} u\in\widetilde{\mathscr{Z}}(r)\right\}
		= \{ \widetilde{W}^HU|U\in \mathscr{\widetilde{Z}}(r) \}. \label{b3}
		\end{align}
	\end{enumerate}

	For the sets $\widetilde{\mathscr{X}}(r)$, $\widetilde{\mathscr{Z}}(r)$, and $\widetilde{\mathscr{U}}(r)$ describing the inverse mappings of three simple component mappings, the set $\widetilde{\mathscr{U}}(r)$ is to take the inverse Fourier transform of $U\in \mathscr{\widetilde{Z}}(r)$ and the set $\widetilde{\mathscr{Z}}(r)$ is to take the square root of $|U|^2$, assign arbitrary phase, and preserve the symmetry of the $U$ for any $|U|^2\in\widetilde{\mathscr{X}}(r)$.
	While  the set $\widetilde{\mathscr{X}}(r)$ is a convex polytope and more detailed properties are given in the following proposition.
	\begin{prop}
		\label{prop7}
		\begin{enumerate}[i).]
			\item When $N \geq 2n$,  $\widetilde{\mathscr{X}}(r)$ is a convex polytope with  at least one element and its dimension is less than or equal to $N/2-n+1$.
			\item When $n\leq N <2n$,
			$\widetilde{\mathscr{X}}(r)$ is singleton.
			In particular, when $N=n$, the unique element is
			$R/\sqrt{n}$, 
			where $R=[R_0,R_1,\cdots,R_{n-1}]^T$ and $\{R_k,0\leq k \leq  n-1 \}$ are the Fourier transform coefficients of $\{r_i,0\leq i \leq n-1 \}$.
		\end{enumerate}
	\end{prop}	
	The explicit expression of $\widetilde{\mathscr{X}}(r)$ for the singleton case is given in the proof of Proposition 1 in Appendix.
	
	\subsection{Comparison with the FDIE in  \cite{Jansson2004}}
	
	It is worth to make detailed comparisons with the FDIE proposed in  \cite{Jansson2004}. To this goal, we first have a brief review accordingly. 
	For the periodic input \dref{eq:assoninicond}, an input design problem is proposed in  (4.33)--(4.34) of Chapter 4  in \cite{Jansson2004} for any finite sample size $N$. That is to  minimize the root mean square (RMS) subject to a frequency constraint
	\begin{subequations}
		\label{jason}
		\begin{align}
		& \underset{\alpha_k,k=0,\cdots,N-1}{\rm minimize}~~\sum_{k=0}
		^{N-1} \alpha_k \label{op1}\\
		&\mbox{subject to} ~~\frac{\sigma^2}{N}\Gamma^H (e^{j\omega})(\Phi^T\Phi)^{-1}\Gamma (e^{j\omega})
		\leq \frac{b(\omega)}{|F(e^{j\omega})|^2},\label{op1c}\\
		&\hspace{18mm}\alpha_k=|U_k|^2,
		\end{align}
	\end{subequations}
	where  
	$F(q)$ is a known stable transfer function,
	$\Gamma (q^{-1})=[q^{-1},\cdots,q^{-n}]^T$,
	and $\Phi^T\Phi$ has the form \dref{phiphi}.

	The inequality  constraint  \dref{op1c} is non-convex in $|U_k|$ and so the problem \dref{jason} is not tractable.
	As a result, two methods are introduced to transform 	the problem into a convex problem in \cite{Jansson2004}. 
	The first one is a solution based on geometric programming, which requires that
	$
	m=n
	$,
	where   $m$ is the number of the nonzero DFT coefficients and $n$ is  the number of the estimated parameters (See (4.38) and (4.39) on page 103 of \cite{Jansson2004}).
	The second one is a solution based on the linear matrix inequality (LMI), where $m$ can be larger than $n$ (See (4.41) on page 105 of \cite{Jansson2004}).
	After the optimal squared magnitude $|U_k|^2$ corresponding to the chosen $m$ nonzero spectral lines are obtained,
	the optimal input can be found by the route consisting of two steps: 1) take the square-root of $|U_k|^2$ with a proper phase setting, 2) take the Fourier transform \dref{ifft}, which is briefly denoted by $|U|^2\rightarrow U\rightarrow u$ in the following.
	
	Here, we would like to highlight the differences between the problem \dref{jason} and our problem \dref{op}:
	\begin{itemize}
		\item 
		
		the optimization variables of \dref{jason} are the DFT coefficients $U_k$  or its squared magnitude $|U_k|^2$ of the input rather than its correlation sequence $r$.
		After the optimal $|U_k|^2$ is found, then the two-step method above gives the optimal input. In other words, the method proposed in Chapter 4 of \cite{Jansson2004} does not determine $r$ first, i.e., it does not formulate the input design problem with $r$ as the optimization variable.

		\item  the route $|U|^2\rightarrow U\rightarrow u$ provides a route to find the optimal input $u$ by the optimal squared magnitude $|U|^2$ in terms of the DFT. 
		Here, we derive a route from $r$ to $u$  by the FDIE \dref{b1}--\dref{b3}: $r\rightarrow|U|^2\rightarrow U\rightarrow u$ based on the DFT, which characterizes all the $u$'s satisfies $f(u)=r$. Moreover, the set $\widetilde{\mathscr{X}}(r)$ given in \dref{b1} consists of all the $|U|^2$s related to the optimal $r$
		and its property is clearly characterized in Proposition \ref{prop7}, and it will also be shown in Theorem \ref{thm2} below that the FDIE \dref{b1}--\dref{b3}: $r\rightarrow|U|^2\rightarrow U\rightarrow u$ is equivalent to the TDIE in terms of the TDE \dref{tdd}, namely, both of them characterize the set $f^{-1}(r)$, and the corresponding computational complexities of these two inverse embeddings are almost the same.
		
	\end{itemize}
	

	\section{Graph Induced Embeddings: A Unified Perspective}\label{secgraph}
	This section first investigates the relation between the TDE and the FDE  as well as their inverse embeddings  and  then applies the graph signal processing~(\cite{Sandryhaila2013}) for a ring graph  to interpret them in a unified perspective
	under the periodic assumption \dref{eq:assoninicond} on  input sequences.
	
	
	\subsection{Connections between Two  Embeddings}
	Let us rewrite the FDE \dref{autocorfft} and the TDE \dref{tdd} as
	\begin{subequations}
		\label{dd}
		\begin{align}	&U=\widetilde{W}u,~|U|^2 = [|U_0|^2,\cdots,|U_{N-1}|^2]^T,~r=\widetilde{S}|U|^2,	\label{fd}\\
		&z=W^Tu,~z^2=[z_0^2,z_1^2\cdots,z_{N-1}^2]^T,~r = Sz^2. \label{td}
		\end{align}
	\end{subequations}
	Since the element-wise quadratic mappings \dref{dd} in between two linear transforms are the same, we denote the FDE and the  TDE by $(\widetilde{W},\widetilde{S})$ and $(W^T,S)$ for convenience, respectively.
	Their connections are established in the following proposition.
	\begin{prop} 
		\label{thm4}
		For  the FDE $(\widetilde{W},\widetilde{S})$ and the TDE $(W^T,S)$, there holds that
		\begin{align}
		W^T=\Lambda \widetilde{W},~~S=(\widetilde{S} + \overline{\widetilde{S}})/2, 
		\end{align}
		where $\Lambda$ is a unitary matrix.
		In particular, for even $N$, we have
		\begin{align*}
		\Lambda=\left[
		\begin{array}{cccccccc}
		1&0&\cdots&0&0&0&\cdots&0\\
		0&\frac{1}{\sqrt{2}}&&&&&&\frac{1}{\sqrt{2}}\\
		\vdots&&\ddots&&&&\iddots   &\\
		0&&&\frac{1}{\sqrt{2}}&0&\frac{1}{\sqrt{2}}&&\\
		0&0&\cdots&0&1&0&\cdots&0\\
		0&&&-\frac{j}{\sqrt{2}}&0&\frac{j}{\sqrt{2}}&&\\
		\vdots&&\iddots& && &\ddots&\\
		0&-\frac{j}{\sqrt{2}}&\cdots&&&&&\frac{j}{\sqrt{2}}\\
		\end{array}
		\right]
		\end{align*}
		and	for odd $N$, we have
		\begin{align*}
		\Lambda=\left[
		\begin{array}{ccccccc}
		1&0&\cdots&0&0&\cdots&0\\
		0&\frac{1}{\sqrt{2}}&&&&&\frac{1}{\sqrt{2}}\\
		\vdots&&\ddots&&&\iddots   &\\
		0&&&\frac{1}{\sqrt{2}}&\frac{1}{\sqrt{2}}&&\\
		0&&&-\frac{j}{\sqrt{2}}&\frac{j}{\sqrt{2}}&&\\
		\vdots&&\iddots& & &\ddots&\\
		0&-\frac{j}{\sqrt{2}}&&&&&\frac{j}{\sqrt{2}}\\
		\end{array}
		\right].
		\end{align*}
		In addition, the vectors $z$ and $U$ satisfy
		$	z=\Lambda U.$
	\end{prop}
	The following theorem shows that the FDIE  \dref{b1}--\dref{b3} based on the composition \dref{tdf} is equivalent to the TDIE.
	\begin{thm}
		\label{thm2}
		Given any $r\in \mathscr{F}$, the set $\widetilde{\mathscr{U}}(r)$ determined by \dref{b1}--\dref{b3} is the same as the set $\mathscr{U}(r)$ produced by the TDIE and both of them are equal to the set $f^{-1}(r)$.
	\end{thm}
	{\bf Example 1.} We use a simple case $N=4$ and $n$ being one of $1,2,3,4$ to explicitly illustrates how the two inverse embeddings are related with each other.
	
	Given an $r\in\mathscr{F}$, suppose that 
	$$|U|^2=\big[|U_0|^2,|U_1|^2,|U_2|^2,|U_1|^2 \big]\in\widetilde{\mathscr{X}}(r).$$
	Then for a given $0\leq \beta < 2\pi$, define
	$$U=\big[|U_0|,|U_1|e^{j\beta},|U_2|,|U_1|e^{-j\beta}\big]
	\in \widetilde{\mathscr{Z}}(r),$$
	which yields the element $\widetilde{u}=\widetilde{W}^HU \in \widetilde{\mathscr{U}}(r)$.
	In the following, we show how the element $\widetilde{u}$ can also be generated by the TDIE.
	
	For the given vector $|U|^2$ and the chosen $\beta$, define
	\begin{align*}
	z^2=\big[|U_0|^2,2(\cos(\beta))^2|U_1|^2,|U_2|^2,2(\sin(\beta))^2|U_1|^2 \big]
	\end{align*}
	and choose
	\begin{align*}
	z&= \big[|U_0|,\sqrt{2}\cos(\beta)|U_1|,|U_2|,\sqrt{2}\sin(\beta)|U_1| \big],~~
	u=Wz
	\end{align*}
	It follows that
	\begingroup
	\allowdisplaybreaks
	\begin{align*}
	\widetilde{u}&=\widetilde{W}^HU
	=W\Lambda U\\
	&=W
	\left[
	\begin{array}{cccc}
	1&0&0&0\\
	0&\frac{1}{\sqrt{2}}&0&\frac{1}{\sqrt{2}}\\
	0&0&1&0\\
	0&-\frac{j}{\sqrt{2}}&0&\frac{j}{\sqrt{2}}\\
	\end{array}
	\right]
	\left[
	\begin{array}{l}
	|U_0|\\
	|U_1|e^{j\beta}\\
	|U_2|\\
	|U_1|e^{-j\beta}\\
	\end{array}
	\right]
	=Wz=u.
	\end{align*} 
	\endgroup
	Conversely, suppose that
	$
	z^2=\big[z_0^2,z_1^2,z_2^2,z_3^2 \big]$ is chosen from the set
	$\mathscr{X}(r)\!=\!\Big\{x\big|Sx=r,\sum_{i=0}^3x_i=\mathcal{C},x_i\geq 0,0\leq i\leq 3\Big\},$
	where $S$ is defined by \dref{S}.
	Choose the element $z=[\sqrt{z_0^2},-\sqrt{z_1^2},-\sqrt{z_2^2},\sqrt{z_3^2}]^T$ from the set $\Big\{
	[\pm\sqrt{z_0^2},\pm\sqrt{z_1^2},\pm\sqrt{z_2^2},\pm\sqrt{z_3^2}]^T
	\Big\}$ including $2^4=16$ elements 
	 for the given $z^2$
	and let $u=Wz$.
Define
	\begin{align*}
	|U|^2&=\left[  z_0^2,\frac{z_1^2+z_3^2}{2},z_2^2,\frac{z_1^2+z_3^2}{2}\right] \in\widetilde{\mathscr{X}}(r),\\
	|U|&=\left[ |z_0|,\sqrt{\frac{z_1^2+z_3^2}{2}}e^{j\beta},-|z_2|,\sqrt{\frac{z_1^2+z_3^2}{2}}e^{-j\beta}\right]
	\in\widetilde{\mathscr{Z}}(r),
	\end{align*}
	where $\beta = \pi+ \arctan(|z_3|/|z_1|)$.
	Then the vector $\widetilde{u} = \widetilde{W}^HU$ is an element of $\mathscr{\widetilde{U}}(r)$.
	We have
	\begin{align*}
	u&=Wz=\widetilde{W}^H\Lambda^Hz\\
	&=\widetilde{W}^H
	\left[
	\begin{array}{cccc}
	1&0&0&0\\
	0&\frac{1}{\sqrt{2}}&0&-\frac{j}{\sqrt{2}}\\
	0&0&1&0\\
	0&\frac{1}{\sqrt{2}}&0&\frac{j}{\sqrt{2}}\\
	\end{array}
	\right]
	\left[
	\begin{array}{r}
	|z_0|\\
	-|z_1|\\
	-|z_2|\\
	|z_3|\\
	\end{array}
	\right]
	=\widetilde{W}^H U = \widetilde{u}.
	\end{align*} 
	This illustrates the implication of Proposition \ref{thm4} and Theorem \ref{thm2}. $\square$
	
	
	
	\subsection{Graph Interpretation}
\begin{figure}[H]
		\includegraphics[scale=0.47]{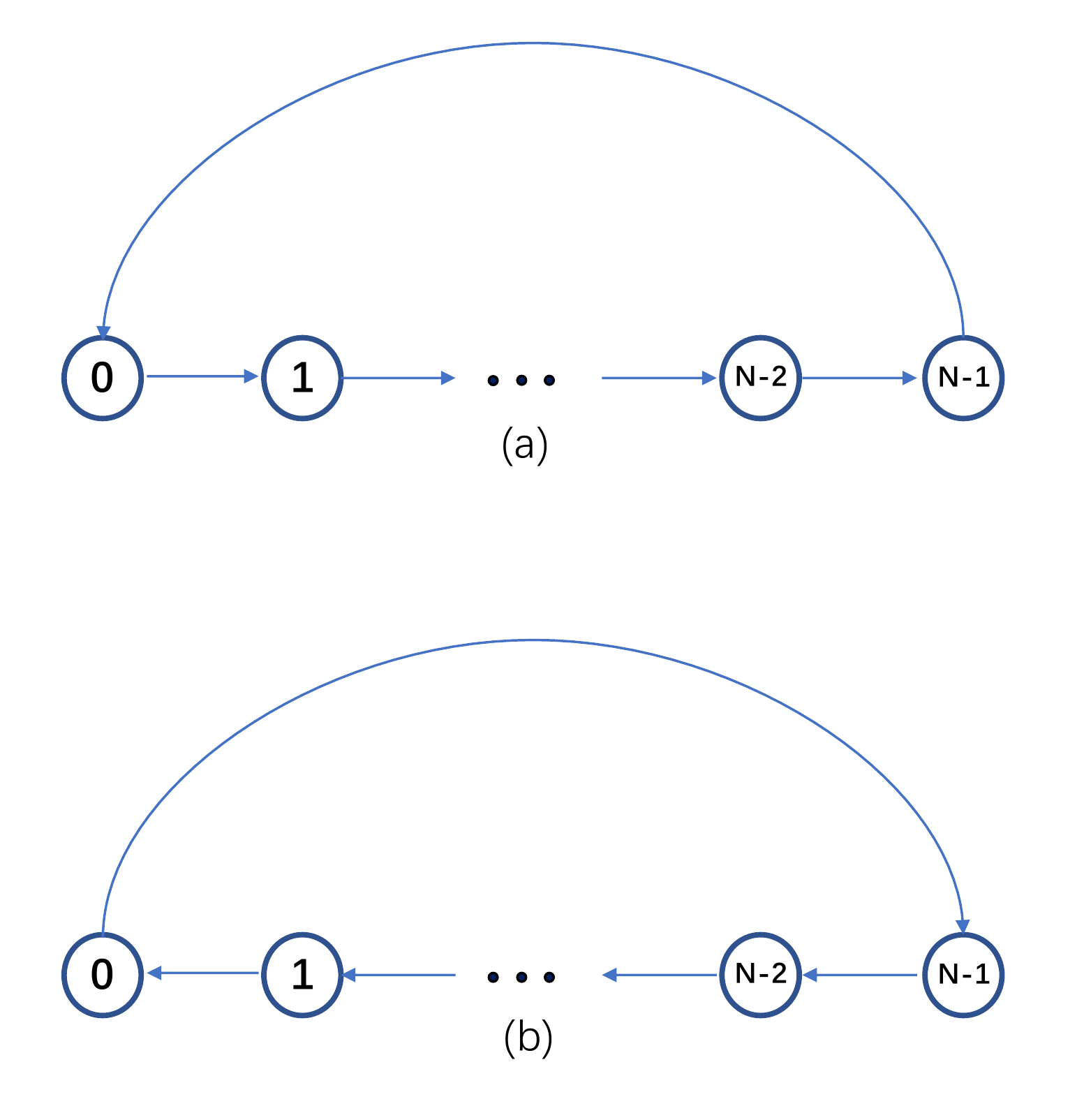}
		\caption{Cyclic graph representation for an $N$-periodic discrete time series. (a) Causal time series. (b) Anti-causal time series. }
		\label{fig11}
	\end{figure}

	Firstly, the FDE \eqref{fd} can be interpreted by a directed cycle graph of the input sequence $\{u_{1-n},\cdots,u_0,\cdots,u_{N-1}\}$ under the periodic assumption \dref{eq:assoninicond}.
	Define a directed cycle graph $\mathscr G\triangleq (\mathscr V, \mathscr E)$ consisting of $N$ nodes in Fig. \ref{fig11}(a), where the set of nodes are $\mathscr V\triangleq\{0,\ldots, N-1\}$,
	the edge set $\mathcal E$ consists of all the  directed edges from each node to its next node with weight $1$.
	The direction of the edges reflects the causality of the time series. 
	We define a graph signal as a mapping from $\mathscr V$ to $\mathbb{R}$, which aligns with the circular input template $u_0,\ldots, u_{N-1}$ for the node set as follows:
	$  i\in\mathscr{V} \mapsto u_i.$
	The cyclic pattern of $\mathscr G$ reflects the periodicity of the input sequence.
	Let $A$ be the adjacent matrix of 
	the graph $\mathscr G$:
	\begin{align}
	A\triangleq 
	\begin{bmatrix}
	0 & \ldots & 0 &1\\
	1 & \ldots & 0 &0\\
	\vdots & \ddots & \vdots & \vdots\\
	0 &\ldots & 1 &0
	\end{bmatrix}. 
	\label{shiftm}
	\end{align}
	Note that $A$ elementwise shifts a signal $u$ forward in a cyclic manner, i.e., 
	$A[u_0,\ldots,u_{N-1}]^T=[u_{N-1},u_0,\ldots, u_{N-2}]^T$. Therefore it is a linear system of the unit delay (also known as the forward shift).  
	
	By Lemma \ref{thma4} in Appendix B, the eigenvalues and the associated unit eigenvectors of
	$A$ are
	$
	\{e^{j\varpi i},~i=0,\ldots, N-1$\}
	and 
	$$
	\left\{\frac{1}{\sqrt{N}}
	[1, e^{-j\varpi i}, \ldots, e^{-j\varpi(N-1)i}]^T,i=0,\ldots, N-1\right\}
	$$
	with $\varpi=2\pi/N$.
	Then it can be seen that
	\begin{itemize}
		\item  the mapping $\widetilde{h}_3(\cdot)$ in \eqref{mf4} corresponds to the Fourier transform of $u$ on $ {\mathscr G}$  (all the rows of $\widetilde{W}$ are exactly the eigenvectors of $A$); 
		
		\item  $\widetilde{h}_1(\cdot)$ in \eqref{mf2} is an expression of the autocovariance of $u$
		in terms of the spectrum $\widetilde{h}_2(\widetilde{h}_3(u))$ of $u$ (the matrix $\widetilde{S}$ consists of all the eigenvalues of $A$).

	\end{itemize}
	

	Secondly, the TDE \dref{td} can be interpreted by a directed cycle graph of the input sequence $\{u_{1-n},\cdots,u_0,\cdots,u_{N-1}\}$ under the periodic assumption \dref{eq:assoninicond}.
	Define the reserve graph ${\mathscr G}'=(\mathscr V,  {\mathscr E}')$ of $\mathscr G$ with the adjacent matrix $A^T$ in Fig. \ref{fig11}(b), where $ {\mathscr E}'$ is a set of directed edges  from
	each node to its last node with weight $1$,
		reflecting  the anti-causality of the time series.
The adjacent matrix elementwise shifts a signal $u$ backward in a cyclic manner, i.e.,
	$A^T[u_0,\ldots,u_{N-1}]^T=[u_1,\ldots, u_{N-1}, u_{0}]^T$. 
	Therefore it is a linear system of the unit advance (also known as backward shift).
	
	\begin{figure}[H]
		\includegraphics[scale=0.47]{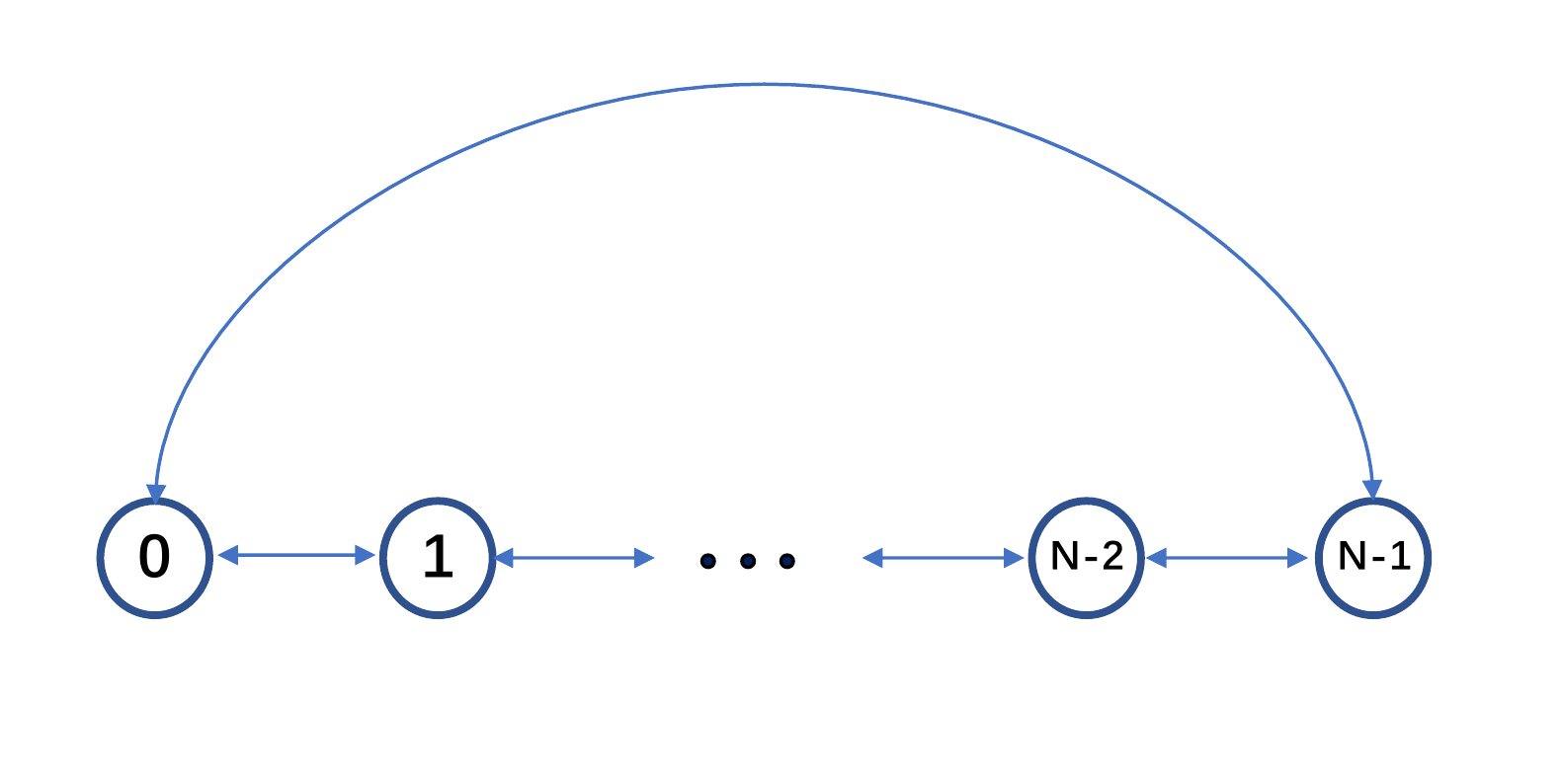}
		\caption{The mirror graph of  an $N$-periodic casual discrete time series. }
		\label{fig12}
	\end{figure}
	Combining $\mathscr G$ and $ {\mathscr G}'$ together, we define the mirror of $\mathscr G$ as $\widetilde {\mathscr G}=(\mathscr V, {\mathscr E}'\cup \mathscr E)$, e.g. Fig.2, whose adjacent matrix is given by 
	$\frac{1}{2}(A+A^T)$ and whose edge set ${\mathscr E}'\cup \mathscr E$ consists of the directed edges from each node to its last and next node with identical weight $1/2$. 
	By Lemma \ref{thma4} in Appendix B, the eigenvalues  of $\frac{1}{2}(A+A^T)$ are 
	$\{\cos(\varpi i), i=0,\ldots, N-1\}$
	and the associated unit eigenvectors are
	$\Big\{\frac{\xi_0}{\sqrt{2}},\xi_1,\cdots,\xi_{\frac{N-2}{2}},\frac{\xi_{\frac{N}{2}}}{\sqrt{2}},\zeta_{\frac{N-2}{2}},\cdots,\zeta_1\Big\}\Big/\sqrt{\frac{2}{N}}$ for even $N$
	and
	$\Big\{\frac{\xi_0}{\sqrt{2}},\xi_1,\cdots,\xi_{\frac{N-1}{2}},\zeta_{\frac{N-1}{2}},\cdots,\zeta_1\Big\}\Big/\sqrt{\frac{2}{N}}$  for odd $N$, where 
	$\xi_i= [1,\cos(i\varpi),\cdots,\cos((N\!\!-\!1)i\varpi)]^T,$~
	$\zeta_i=[0,\sin(i\varpi),\cdots,\sin((N\!\!-\!1)i\varpi)]^T
	$
	for $i\geq 0$.
	Then it can be seen that
	\begin{itemize}
		\item  the mapping $h_3(\cdot)$ in \eqref{md4} corresponds to the graph Fourier transform of $u$ on $\widetilde {\mathscr G}$ \citep{Sandryhaila2013} (all the rows of $W^T$ are exactly all the eigenvectors of $\frac{1}{2}(A+A^T)$); 
		\item $h_1(\cdot)$ in \eqref{md2} is an expression of the autocovariance of $u$
		in terms of the spectrum $h_2(h_3(u))$ of $u$ (the matrix $S$ consists of all the eigenvalues of $\frac{1}{2}(A+A^T)$).

	\end{itemize}

	As a result, the two kinds of  embeddings have been unified from this graph signal processing perspective, which are fully characterized by the eigenvectors and eigenvalues of the matrices $A$ and $\frac{1}{2}(A+A^T)$, respectively.
	
	\subsection{More Embeddings by Graph Diffusions}
	
	Interestingly, it is in fact possible to obtain more embedding from the graph signal processing perspective.  To state this result, we start from an observation. Define a directed graph $\widetilde {\mathscr G}(\gamma)=(\mathscr V, \widetilde{\mathscr E}(\gamma))$ with the adjacency matrix $\widetilde{A}(\gamma)=\gamma A + (1-\gamma)A^T$ for any $\gamma\in\mathbb{C}$, where each node $i$ in $\mathscr V$ is connected to its preceding node ${i-1}$ with the weight $\gamma$ and to its following node ${i+1}$ with the weight $1-\gamma$.
	It follows that $\widetilde{A}(\gamma)[u_0,\ldots,u_{N-1}]^T=\gamma[u_{N-1},u_0,\ldots, u_{N-2}]^T+(1-\gamma)[u_1,\ldots, u_{N-1}, u_{0}]^T$, which is a weighted sum of the forward shift and backward shift.
	Clearly, the two kinds of embeddings above are special cases with $\gamma =1$ and $\gamma =1/2$, respectively.
	

	Then it is natural to raise one question: whether or not the graph $\widetilde {\mathscr G}(\gamma)$ for any $\gamma \in \mathbb{C}$ corresponds to one embedding for the quadratic mapping \dref{eq:defmap}.
	Note that $r_i=u^TA^i u$ and $r_i=u^T(A^T)^iu$ for $i=0,\cdots,n-1$.
	Thus, we have $r_i=u^T(\gamma A^i + (1-\gamma)(A^T)^i)u$ with $i=0,\cdots,n-1$ holds for all $\gamma\in\mathbb{C}$.
	Actually, the key for this question lies in that whether or not $A$ and $A^T$ are simultaneously diagonalizable, and if 
	$A$ and $A^T$ are simultaneously diagonalizable, then the answer is affirmative.
	
	By Lemma \ref{thma4} in Appendix B, the fact that both $A$ and $A^T$ are circular means that they are simultaneously diagonalizable by the matrix $\widetilde{W}$, i.e., 
	\begin{align*}
	A  &= \widetilde{W}\diag([1,e^{-j(N-1)\varpi},\cdots,e^{-j(N-1)^2\varpi)}])\widetilde{W}^H,\\
	A^T  &= \widetilde{W}\diag([1,e^{-j \varpi},\cdots,e^{-j(N-1)\varpi}])\widetilde{W}^H,
	\end{align*}
	which implies that
	\begin{align*}
	A  &= \widetilde{W}^H\diag([1,e^{j(N-1)\varpi},\cdots,e^{j(N-1)^2\varpi)}])\widetilde{W},\\
	A^T  &= \widetilde{W}^H\diag([1,e^{j \varpi},\cdots,e^{j(N-1)\varpi}])\widetilde{W},
	\end{align*}
	since both $A$ and $A^T$ are real.
	It follows that for $i=0,\cdots,n-1$
	\begin{align*}
	r_i &= 
	u^T(\gamma A^i + (1-\gamma) (A^T)^i)^Tu\\
	&=
	u^T
	\widetilde{W}^H
	\Big(\gamma \diag([1,e^{j(N-1)i\varpi},\cdots,e^{j(N-1)^2i\varpi)}])\\
	&\hspace{5mm}+(1-\gamma)
	\diag([1,e^{ji\varpi},\cdots,e^{j(N-1)i\varpi}])
	\Big)
	\widetilde{W} u\\
	&=\sum_{k=0}^{N-1}
	(\gamma e^{j(N-1)ki\varpi}  + (1-\gamma) e^{jki\varpi})
	|U_k|^2\\
	&=\sum_{k=0}^{N-1}
	(\gamma e^{-jki\varpi}  + (1-\gamma) e^{jki\varpi})
	|U_k|^2.
	\end{align*}
	Define the matrix $\widetilde{S}(\gamma)$ of size $n\times N$ with its $(i,j)$-element being
	$\gamma e^{-jki\varpi}  + (1-\gamma) e^{jki\varpi}$.
	Then the embedding of the mapping \dref{eq:defmap} can also be expressed by
	\begin{subequations}
		\label{gembed}
		\begin{align}
		r=\widetilde{S}(\gamma)|U|^2&=\widetilde{h}^\gamma_1(\widetilde{h}_2(\widetilde{h}_3(u))) ~~\mbox{with} \label{gembed1}\\
		\widetilde{h}_1^\gamma(|U|^2)&=\widetilde{S}(\gamma)|U|^2 \label{gembed2}\\
		\widetilde{h}_2(U)&=[|U_1|^2,\cdots,|U_{N-1}|^2]^T \label{gembed3}\\
		\widetilde{h}_3(u) &=U= \widetilde{W} u, \label{gembed4}
		\end{align}
	\end{subequations} and is called the \textit{graph induced embedding} (GIE) of the mapping \dref{eq:defmap}. Since the GIE is the same as the FDE except for the mapping $\widetilde{h}_1^\gamma(\cdot)$, 
	the image sets of the mappings $\widetilde{h}_3$ and $\widetilde{h}_2$ given in \dref{gembed4} and \dref{gembed3} are the same as $\mathscr{Z}$ and  $\mathscr{X} $ given in \dref{image1} and \dref{image2}, respectively.
	
	Similarly to the FDIE route \dref{b1}--\dref{b3}, the inverse embedding of the GIE \dref{gembed}, called the \textit{graph induced inverse embedding} (GIIE),  can be done in the following procedure:
	\begin{enumerate}[i).]
		\item finding the inverse image of $\widetilde{h}_1^\gamma(\cdot)$ for $r\in \mathscr{F}$:
		\begin{align}
		\nonumber
		\hspace{-6mm}
		\widetilde{\mathscr{X}}^\gamma(r)
		\nonumber
		=&\Big\{ |U|^2\Big |
		\widetilde{S}(\gamma)|U|^2 = r,~|U_k|^2\geq 0,0\leq k \leq N-1,\\
		&~~|U_k|^2=|U_{N-k}|^2, 1\leq k \leq N \!-\!1\Big\};
		\label{c1}
		\end{align}
		\item finding the inverse image of $\widetilde{h}_2(\cdot)$ for $|U|^2\in\widetilde{\mathscr{X}}^\gamma(r)$:
		\begin{align}
		\hspace{-8mm}
		\widetilde{\mathscr{Z}}^\gamma(r)
		=\left\{ \scriptsize
		\begin{array}{l}
		\Big\{U\Big|
		U_0=\pm\sqrt{|U_0|^2},~U_{N/2}=\pm\sqrt{|U_{N//2}|^2},\\
		\hspace{6mm} U_k=\sqrt{|U_k|^2}e^{j\beta_k}, ~U_{N-k}=\overline{U_k},\\
		\hspace{6mm} 0\leq \beta_k <2\pi,1\leq k \leq N/2-1
		\Big\}
		~~\mbox{for even } N \\
		\Big\{U\Big|
		U_0=\pm\sqrt{|U_0|^2},U_{N-k}=\overline{U_k} \\
		\hspace{6mm}U_k=\sqrt{|U_k|^2}e^{j\beta_k},
		~,0\leq \beta_k <2\pi\\
		\hspace{6mm} 1\leq k \leq (N-1)/2
		\Big\} ~~ \mbox{for odd } N;
		\end{array}
		\right.
		\label{c2}
		\end{align}
		
		\item finding the inverse image of $\widetilde{h}_3(\cdot)$ for $U\in \widetilde{\mathscr{Z}}^\gamma(r)$:
		\begin{align}
		\hspace{-4mm}\widetilde{\mathscr{U}}^\gamma(r)
		= \{ \widetilde{W}^HU|U\in \mathscr{\widetilde{Z}}^\gamma(r) \}. \label{c3}
		\end{align}
	\end{enumerate}
	
	More formally, we have the following result on the GIE and the GIIE.	
	
	\begin{thm}
		\label{thm3}
		\begin{enumerate}[1)]
			\item All the pairs $(\widetilde{W},\widetilde{S}(\gamma))$ with $\gamma\in \mathbb{C}$ are the embeddings of the mapping \dref{eq:defmap}.
			\item The set $\widetilde{\mathscr{U}}^\gamma(r)$ produced by the GIIE \dref{c1}--\dref{c3} is the same as the  set $\widetilde{\mathscr{U}}(r)$ produced by the FDIE \dref{b1}--\dref{b3}. 
		\end{enumerate}
	\end{thm}
	
It is worth to note that in contrast with the TDE \dref{tdd}
		and the FDE \dref{tdf}, the GIE \eqref{gembed} with $\gamma\in \mathbb C$ has an extra design freedom corresponding to the choice of $\gamma$. Clearly, how to make use of this extra design freedom is an interesting problem and will be studied in details in the future. In what follows, to shed some light on this problem, 
		we study the case $\gamma=1/2$, which corresponds to the directed graph in which 
		each node $i$ in $\mathscr V$ is connected to its
		neighboring nodes
		${i-1}$ and ${i+1}$ with the identical weight $1/2$. In this case, the matrix $\widetilde{S}(1/2)$ reduces to a real matrix 
	\begin{align*}
	\begin{bmatrix}
	1&1&\cdots&1\\
	1&\cos(\varpi)&\cdots&\cos((N-1)\varpi)\\
	\vdots & \vdots & \ddots & \vdots\\
	1&\cos(\varpi(n-1))&\cdots&\cos((N-1)(n-1)\varpi)
	\end{bmatrix}
	\end{align*}
	and its $k$-th  and  $(N-k)$-th columns  are identical since $\cos(ik\varpi)=\cos(i(N-k))$ for $i=0,1\cdots,n-1$ and $k=0,1\cdots,N-1$.
	Also, $\widetilde{S}(1/2)$  is the unique real matrix among all $\gamma\in\mathbb{C}$.
	This property of the matrix $\widetilde{S}(1/2))$ leads to more embeddings besides the embedding $(\widetilde{W},\widetilde{S}(1/2))$ for the mapping \dref{eq:defmap}.
	For even $N$, define the set $\Lambda_Q$ consisting of all the unitary matrices having the form of
	\begin{align*}
	\left[
	\begin{array}{cccccccc}
	1&0&\cdots&0&0&0&\cdots&0\\
	0&q_{11}^{(1)}&&&&&&q_{12}^{(1)}\\
	\vdots&&\ddots&&&&\iddots   &\\
	0&&&q_{11}^{((N-2)/2)}&0&q_{12}^{((N-2)/2)}&&\\
	0&0&\cdots&0&1&0&\cdots&0\\
	0&&&q_{21}^{((N-2)/2)}&0&q_{22}^{((N-2)/2)}&&\\
	\vdots&&\iddots& && &\ddots&\\
	0&q_{21}^{(1)}&\cdots&&&&\cdots&q_{22}^{(1)}\\
	\end{array}
	\right],
	\end{align*}
	where 
	$Q_i=\begin{bmatrix}
	q_{11}^{(i)}&q_{12}^{(i)}\\
	q_{21}^{(i)}&q_{22}^{(i)}
	\end{bmatrix}
	$
	for $i=1,\cdots,(N-2)/2$ are arbitrary unitary matrices of size $2\times 2$.
	Similarly, we define the set $\Lambda_Q$ for odd $N$.
	

	Then  we have the following result on the embeddings of the mapping \dref{eq:defmap} corresponding to the directed graph with the weight $\gamma=1/2$.
	
	\begin{thm}
		\label{thm5}
		When $\gamma=1/2$, all the pairs $(\widetilde{\Lambda}\widetilde{W},\widetilde{S}(1/2))$ with $\widetilde{\Lambda} \in \Lambda_Q$ are the embeddings of the mapping \dref{eq:defmap}.
		
	\end{thm}
	
	\begin{rem}
		Clearly,		the FDE \dref{tdf} corresponds to the pair  $(\widetilde{W},\widetilde{S}(1))$, while 
		the TDE \dref{tdd} corresponds to the pair $(\Lambda\widetilde{W},\widetilde{S}(1/2))$, where $\Lambda$ is defined in Proposition \ref{thm4}. 
	\end{rem}
	
	Note that the pair $(\Lambda\widetilde{W},\widetilde{S}(1/2))$ is a real embedding. Then we are wondering whether or not there are more real embeddings besides $(\Lambda\widetilde{W},\widetilde{S}(1/2))$ among all the embeddings $(\widetilde{\Lambda}\widetilde{W},\widetilde{S}(1/2))$ with $\widetilde{\Lambda} \in \Lambda_Q$.
	If true, how are these real embeddings related to each other?
	
	The existence of real embeddings is because the $k$-th row and the $(N\!-\!k)$-row of $\widetilde{W}$ are complex conjugate.
	Thus, the matrix $\widetilde{\Lambda}\widetilde{W}$ is real if each $Q_i$ in $\widetilde{\Lambda}$ chooses one from the following eight unitary candidates 
	\begin{align*}
	& \begin{bmatrix}
	\frac1{\sqrt{2}}&\frac1{\sqrt{2}}\\
	-\frac{j}{\sqrt{2}}&\frac{j}{\sqrt{2}}
	\end{bmatrix},
	\begin{bmatrix}
	-\frac1{\sqrt{2}}&-\frac1{\sqrt{2}}\\
	-\frac{j}{\sqrt{2}}&\frac{j}{\sqrt{2}}
	\end{bmatrix},
	\begin{bmatrix}
	\frac1{\sqrt{2}}&\frac1{\sqrt{2}}\\
	\frac{j}{\sqrt{2}}&-\frac{j}{\sqrt{2}}
	\end{bmatrix},
	\begin{bmatrix}
	-\frac1{\sqrt{2}}&-\frac1{\sqrt{2}}\\
	\frac{j}{\sqrt{2}}&-\frac{j}{\sqrt{2}}
	\end{bmatrix},\\
	&\begin{bmatrix}
	-\frac{j}{\sqrt{2}}&\frac{j}{\sqrt{2}}\\
	\frac1{\sqrt{2}}&\frac1{\sqrt{2}}
	\end{bmatrix},
	\begin{bmatrix}
	-\frac{j}{\sqrt{2}}&\frac{j}{\sqrt{2}}\\
	-\frac1{\sqrt{2}}&-\frac1{\sqrt{2}}
	\end{bmatrix},
	\begin{bmatrix}
	\frac{j}{\sqrt{2}}&-\frac{j}{\sqrt{2}}\\
	\frac1{\sqrt{2}}&\frac1{\sqrt{2}}
	\end{bmatrix},
	\begin{bmatrix}
	\frac{j}{\sqrt{2}}&-\frac{j}{\sqrt{2}}\\
	-\frac1{\sqrt{2}}&-\frac1{\sqrt{2}}
	\end{bmatrix}.
	\end{align*}
	Clearly, each  matrix  above plays the following roles when it is embedded in $\widetilde{\Lambda}$: 1) extract the real part and imaginary part of the corresponding two rows of $\widetilde{W}$; 2) assign a sign (positive or negative) for the two rows; 3) keep or change the order of the two rows.
	Otherwise, if at least one of $Q_i$s is not chosen from the eight matrices, then  $\widetilde{\Lambda}\widetilde{W}$ will involve complex numbers.
	Thus, we have the following conclusion.
	\begin{thm}
		\label{thm6}
		There are $8^{\frac{N-2}{2}}$ real embeddings for even $N$ ($8^{\frac{N-1}{2}}$ real embeddings for odd $N$) among all the embeddings having the form of $(\widetilde{\Lambda}\widetilde{W},\widetilde{S}(1/2))$ with $\widetilde{\Lambda} \in \Lambda_Q$.
		Moreover, the pair $(\Lambda\widetilde{W},\widetilde{S}(1/2))$ is unique up to signs and orders of the rows of $\widetilde{\Lambda}\widetilde{W}$.
	\end{thm}	
	
	\section{Numerical Illustration}
	
	The GIIE procedure \dref{c1}--\dref{c3} characterizes all the inputs $u$ corresponding to a given autocovariance $r$ in the set $\mathscr{F}$.
This section uses a numerical example to verify this procedure.
	
	
	
	The setting of the numerical example is as follows: $N=120,n=50,\mathcal{C}=120,\sigma^2=0.5$ and the TC kernel with the scale hyperparameter and the decaying one equal to $1$ and $0.85$, respectively.

	Firstly, the optimal autocovariance $r^*$ corresponding to the setting above is obtained by using the CVX software package developed in \cite{Grant2016} to solve the convex optimization problem \eqref{or}.
	
	 Secondly, we randomly generate 100 inputs corresponding to $r^*$ and each input is generated by the GIIE \dref{c1}--\dref{c3} in the following way:  given $\gamma=1/2$,
	\begin{itemize}
		\item[1)] obtain a $|U|^2$ in the set \eqref{c1} by using the {\ttfamily fmincon} function in MATLAB to solve the optimization problem 
		\begin{align*}
		&\min_{|U|^2\in\Omega} \big(\widetilde{S}(1/2)|U|^2 - r^*\big)^2\\
		&\Omega=\Big\{|U|^2\Big|\sum_{k=0}^{N-1}|U|^2_k\! =\! \mathcal{C},|U|^2_k \geq 0,k\!=\!0,1,\dots,N\!-\!1,\\
		&\hspace{20mm}|\overline{U_k}|^2=|U_{N-k}|^2,~k=1,\cdots,N\!-\!1 \big\}
		\end{align*}
		with a randomly chosen starting point in $\Omega$.
		\item[2)] obtain a $U$ in the set \eqref{c2} corresponding to the $|U|^2$ produced in 1) by independently and randomly choosing all of the phase parameters $\beta_k,k=1,\cdots,N/2-1$ from the uniform distribution $[0,2\pi)$.
		\item[3)] obtain an input $u$ by taking the inverse Fourier transform of the $U$ obtained in 2).
	\end{itemize}
	
\begin{figure}[h]
			\hspace*{-1em}
		\includegraphics[scale=0.42]{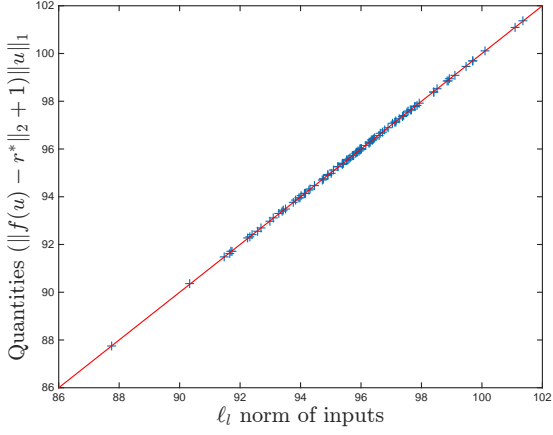}
		\caption{The produced 100 inputs by the GIIE procedure.}
		\label{fig1}
	\end{figure}

	For each input $u$ produced above, it needs to check whether $f(u)=r^*$ or not.
	To show that all of the generated 100 inputs indeed belong to the inverse image set of the quadratic mapping \dref{eq:defmap} corresponding to $r^*$  in a visual form.
	We plot the 100 pairs of the quantities
	$\{\|u\|_1,(\|f(u) - r^*\|_2+1)\|u\|_1\}$
	related to the 100 inputs in Fig. \ref{fig1}, where each blue plus symbol represents one input, the red solid line is the straight line $y=x$,
and $\|\cdot\|_1$ and $\|\cdot\|_2$  are the  $\ell_1 $ and $\ell_2$ norms of a column vector.
If $f(u)=r^*$, then $(\|f(u) - r^*\|_2+1)\|u\|_1=\|u\|_1$.
Therefore, we can claim that all the 100 inputs by the GIIE procedure \dref{c1}--\dref{c3} are located in the inverse image $f^{-1}(r^*)$ if all the 100 pairs are on the line $y=x$ of Fig.  \ref{fig1}.
We see from Fig. \ref{fig1} that all the 100 produced inputs are almost on the straight line $y=x$. This means that the GIIE \dref{c1}--\dref{c3} indeed characterizes the inverse embedding of the quadratic mapping.

	\section{Conclusions}
	\label{con}
	
	This paper took steps forward for the QMIE method proposed for solving the input design of the RLS estimator.
	Firstly, the FDIE of the quadratic mapping was developed for general cases $N\geq n$ based on the well-known FDE of the mapping. 
	Secondly, a clear connection between the FDIE and the TDIE of the quadratic mapping was discovered and these two embeddings actually correspond to two special directed graphs of periodic signals, respectively.
	Lastly, more embeddings corresponding to different directed graphs of periodic signals were found in a unified graph signal processing perspective and the real embedding is unique if ignoring the signs and orders of the rows of the orthogonal matrix.

	\section*{Acknowledgments}
	The authors thank an anonymous reviewer of \cite{Mu2018b} for bringing out the possibility of a new line of research from frequency domain, which inspired us to complete the current work.
	
		This work was supported in part by the National Key R\&D Program of China under Grant 2018YFA0703800,
		the Strategic Priority Research Program of Chinese Academy of Sciences under Grant No. XDA27000000,
		the NSFC under Grant Nos. 61773329 and 11971239, 
		the Thousand Youth Talents Plan funded by the central
		government of China, the Shenzhen Research Projects Ji-20170189 and Ji-20160207 funded by the Shenzhen Science and Technology Innovation Council, the
		Presidential Fund PF. 01.000249 funded by the Chinese
		University of Hong Kong, Shenzhen, 
		the Science, Technology, and Innovation Commission of Shenzhen Municipality under Grant No. ZDSYS20200811143601004,
		Natural Science Foundation of the
		Higher Education Institutions of Jiangsu Province under Grant 21KJA110002, and
		Shenzhen Science and Technology Programs ZDSYS20211021111415025 and JCYJ20210324120011032.

	\appendix
	\textbf{Appendix A}
	\renewcommand{\thesection}{A}
	\setcounter{thm}{0}
	\renewcommand{\thethm}{A\arabic{thm}}
	\setcounter{lem}{0}
	\renewcommand{\thelem}{A\arabic{lem}}
	\setcounter{rem}{0}
	\renewcommand{\therem}{A\arabic{rem}}
	\renewcommand{\thesection}{A}

	The appendix A contains the proofs of the results in the paper.

	\subsection{Proof of Proposition  \ref{prop7}}
	
	Firstly, by definition the set $\widetilde{\mathscr{X}}(r)$ is nonempty if $r\in \mathscr{F}$.
	
	We intend to prove the conclusion by considering even $N$ and odd $N$, respectively.
	
	For the case of even $N$, by using the constraints $\{|U_k|^2=|U_{N-k}|^2, 1\leq k \leq N \!-\!1\}$ of $\widetilde{\mathscr{X}}(r)$,
	we have
	\begingroup
	\allowdisplaybreaks
	\begin{align*}
	\nonumber
	\widetilde{\mathscr{X}}(r)
	&=\Big\{ |U|^2\Big||U_k|^2\geq 0,0\leq k \leq N-1,\\
	\nonumber
	&\hspace{15mm} |U_{N-k}|^2=|U_k|^2, 1\leq k \leq N/2-1,\\
	&\hspace{15mm} \underline{S}\underline{|U|^2} =r\Big\}\\
	\underline{S}
	&\eq\left[
	\begin{array}{ccccc}
	1&1&\cdots&1\\
	1&\cos(\varpi) &\cdots&\cos(\varpi (N/2))\\
	\vdots&\vdots&\ddots&\vdots\\
	1&\cos(\varpi(n-1))&\cdots&\cos(\varpi((n-1)N/2))
	\end{array}
	\right]\!\\
	\underline{|U|^2}
	&\eq\left[
	|U_0|^2,
	2|U_1|^2,
	\cdots,
	2|U_{N/2-1}|^2,
	|U_{N/2}|^2		
	\right]^T\!\!.
	\end{align*}
	\endgroup
	Note that the size of $\underline{S}$ is $n\times (N/2+1)$.
	Thus $\rank(\underline{S})=\min(N/2+1,n)$ by Theorem 2 of \cite{Mu2018b}.
	So, it is clear that $\underline{S}$ is of  full column rank and accordingly the linear equation $\underline{S}\underline{|U|^2} =r$ only has one solution if $N\leq 2(n-1)$.
	Suppose that 
	$$^*\underline{|U|^2}
	=\left[
	^*|U_0|^2,
	^*|U_1|^2,
	\cdots,
	^*|U_{N/2-1}|^2,
	^*|U_{N/2}|^2		
	\right]^T$$ is the unique solution
	and note that $\widetilde{\mathscr{X}}(r)$ contains at least one element.
	Then $\widetilde{\mathscr{X}}(r)$ consists of the unique element 
	\begin{align*}
	\big[
	^*|U_0|^2,
	^*|U_1|^2/2,
	\cdots,
	&^*|U_{N/2-1}|^2/2,
	^*|U_{N/2}|^2,\\
	& ^*|U_{N/2-1}|^2/2,
	\cdots,
	^*|U_1|^2/2
	\big]^T
	\end{align*}
	when $N\leq 2(n-1)$ for even $N$.
	
	On the other hand, 
	the dimension of the null space of the linear equation $\underline{S}\underline{|U|^2} =r$ is $N/2+1-n$ when  $N\geq 2n$ for even $N$.
Let  $^*|U|^2$ be one element in $\widetilde{\mathscr{X}}(r)$.	
This implies that the set
	\begin{align}
	\nonumber
\widetilde{\mathscr{Y}}(r)	&\eq \Big\{ |U|^2\Big |
	\widetilde{S}|U|^2 = r,~|U_k|^2=|U_{N-k}|^2, 1\leq k \leq N \!-\!1\Big\}\\
	&=^*|U|^2\oplus {\rm Range}\{\xi_j,j=n,\cdots,N/2\}
	\end{align}
	with $\xi_j$ being defined by \eqref{xi}, is an affine space of dimension  $N/2-n+1$, where $\oplus$ means the direct sum of two subspaces and ${\rm Range}\{\cdot\}$ means the range of columns.
	Therefore, we have the dimension of $\widetilde{\mathscr{X}}(r)$ is less than or equal to $N/2-n+1$ since $\widetilde{\mathscr{X}}(r)\subset \widetilde{\mathscr{Y}}(r)$.

	Similarly, for the case of odd $N$, we have
	\begingroup\allowdisplaybreaks
	\begin{align*}
	\nonumber
	\widetilde{\mathscr{X}}(r)
	&=\Big\{ |U|^2\Big| |U_k|^2\geq 0,0\leq k \leq N-1\\
	\nonumber
	&\hspace{15mm} |U_{N-k}|^2=|U_k|^2, 1\leq k \leq (N-1)/2\\
	&\hspace{15mm} \underline{S}\underline{|U|^2} =r\Big\}\\
	\underline{S}
	&\eq\left[
	\begin{array}{ccccc}
	1&1&\cdots&1\\
	1&\cos(\varpi)&\cdots&\cos(\varpi((N\!-\!1)/2))\\
	\vdots&\vdots&\ddots&\vdots\\
	1&\cos(\varpi(n-1))&\cdots&\cos(\varpi((n\!-\!1)(N\!-\!1)/2))
	\end{array}
	\right]\!\\
	\underline{|U|^2}
	&\eq\left[
	|U_0|^2,
	2|U_1|^2,
	\cdots,
	2|U_{(N-1)/2}|^2		
	\right]^T\!\!.
	\end{align*}
	\endgroup
	Thus, the matrix $\underline{S}$ is of  full column rank and accordingly
	$\underline{S}\underline{|U|^2} =r$ has one solution if $N\leq 2n-1$.
	Suppose that
	$^*\underline{|U|^2}
	=\left[
	^* |U_0|^2,
	^*|U_1|^2,
	\cdots,
	^*|U_{(N-1)/2}|^2		
	\right]^T$
	is the unique solution of $\underline{S}\underline{|U|^2} =r$.
	Then $\widetilde{\mathscr{X}}(r)$ includes only one element
	\begin{align*}
	\big[
	^*|U_0|^2,
	&^*|U_1|^2/2,
	\cdots,
	^*|U_{(N-1)/2}|^2/2,\\
	&^*|U_{(N-1)/2}|^2/2,
	\cdots,
	^*|U_1|^2/2
	\big]^T.
	\end{align*}
	Meanwhile, $\widetilde{\mathscr{X}}(r)$ is a convex polytope and its  dimension is less than or equal to $(N+1)/2-n$ by a similar statement as for the even $N$ case when  $N\geq 2n+1$ for odd $N$.
	
	Therefore, by combining the results for even  $N$ and odd $N$, we proved that $\widetilde{\mathscr{X}}(r)$ has one element when $N< 2n$.
	
	When $N=n$, by the previous derivation, $\widetilde{\mathscr{X}}(r)$ has only one element and so for proving the vector
	$R=[R_0,R_1,\cdots,R_{n-1}]^T$
	is the unique one, one just needs to show that
	this vector is a solution of the linear equation $\widetilde{S}R/\sqrt{n}=r$,
	which is verified straightforwardly by the inverse Fourier transform  from $R$ to $r$.
	
	This completes the proof.
	
	\subsection{Proof of Proposition  \ref{thm4}}
	The proof is straightforward by comparing the corresponding matrices and is omitted.

	%
	%
	
	\subsection{Proof of Theorem  \ref{thm2}}
	For the sets $\mathscr{X}(r)$, $\mathscr{Z}(r)$, $\mathscr{U}(r)$ involved in the proof, please refer to Theorem 1 of \cite{Mu2018b}.
	
	We first prove the theorem for even $N$.
	
	Let $\widetilde{u}$ be any element of $\mathscr{\widetilde{U}}(r)$.
	Then, we have $f(\widetilde{u})=r$ and accordingly there exists a route to generate $\widetilde{u}$ from $r$ by the way of \dref{b1}--\dref{b3}.
	First of all, based on $\widetilde{u}$ we have the element $\widetilde{U}=\widetilde{W} \widetilde{u}\in \mathscr{\widetilde{Z}}(r)$ denoted by $[\widetilde{U}_0,\cdots,\widetilde{U}_{N-1}]^T$.
	Second of all, let us define
	\begin{align*}
	&|\widetilde{U}|^2=[|\widetilde{U}_0|^2,\cdots,|\widetilde{U}_{N-1}|^2]^T
	\end{align*}
	where $|\widetilde{U}_k|^2=|\widetilde{U}_{N-k}|^2$ for $1\leq k \leq N-1$.
	Thus, we have $\widetilde{S}|\widetilde{U}|^2=r$ and hence $|\widetilde{U}|^2\in\mathscr{\widetilde{X}}(r)$ since $f(\widetilde{u})=r$.
	Now define $z=\Lambda \widetilde{U}	=[z_0,z_1\cdots,z_{N-1}]^T,z^2=[z_0^2,z_1^2\cdots,z_{N-1}^2]^T$,
	where $\Lambda$ is defined in Proposition \ref{thm4}.
	By the relation $z=\Lambda U$ given in  Proposition \ref{thm4}, we have
	\begin{subequations}
		\label{even}
		\begin{align}
		&z_0^2 = |\widetilde{U}_0|^2,~~z_{N/2}^2=|\widetilde{U}_{N/2}|^2\\
		&z_k^2=2\times({\rm Re}(\widetilde{U}_k))^2,~~
		z_{N-k}^2=2\times({\rm Im}(\widetilde{U}_k))^2
		\end{align}
	\end{subequations}
	for $1\leq k \leq N/2-1$, where ${\rm Re}(\cdot)$ and ${\rm Im}(\cdot)$ denote the real part and the imaginary part of a complex number, respectively.
	This yields that $z_k^2+z_{N-k}^2=2|\widetilde{U}_k|^2=2|\widetilde{U}_{N-k}|^2$ for $1\leq k \leq N/2-1$.
	It follows that
	\begingroup
	\allowdisplaybreaks
	\begin{align*}
	Sz^2&=
	\left[
	\begin{array}{ccccc}
	1&1&\cdots&1\\
	1&\cos(\varpi)&\cdots&\cos(\varpi(N-1))\\
	\vdots&\vdots&\ddots&\vdots\\
	1&\cos(\varpi(n-1))&\cdots&\cos(\varpi(n-1)(N-1))
	\end{array}
	\right]\\
	&\hspace{5mm}\times
	\left[
	\begin{array}{c}
	z_0^2\\
	z_1^2\\
	\vdots\\
	z_{N-1}^2	
	\end{array}
	\right]\\
	&=
	\left[
	\begin{array}{ccccc}
	1&1&\cdots&1\\
	1&e^{-j\varpi}&\cdots&e^{-j\varpi(N-1)}\\
	\vdots&\vdots&\ddots&\vdots\\
	1&e^{-j\varpi(n-1)}&\cdots&e^{-j\varpi(n-1)(N-1)}
	\end{array}
	\right]
	\left[
	\begin{array}{c}
	|\widetilde{U}_0|^2\\
	|\widetilde{U}_1|^2\\
	\vdots\\
	|\widetilde{U}_{N-1}|^2	
	\end{array}
	\right]\\
	&=\widetilde{S}|\widetilde{U}|^2=r.
	\end{align*} 
	\endgroup
	This means that $z^2\in\mathscr{X}(r)$
	and also $z\in \mathscr{Z}(r)$.
	Furthermore, we define  $u=Wz$ and  hence $u\in\mathscr{U}(r)$.
	Now, one still needs to show $\widetilde{u}=u$, which is verified by
	\begin{align*}
	u=Wz=(\Lambda \widetilde{W})^Hz=(\Lambda \widetilde{W})^H\Lambda \widetilde{U}=\widetilde{W}^H\widetilde{U}=\widetilde{u}.
	\end{align*}
	Conversely, it also requires to show that, for any element $u\in\mathscr{U}(r)$, there exist  the corresponding elements  $|\widetilde{U}|^2\in\mathscr{\widetilde{X}}(r)$ and
	$\widetilde{U}\in\mathscr{\widetilde{Z}}(r)$ such that $u=\widetilde{W}^H\widetilde{U}$.
	By applying the inverse mapping of \dref{even}, namely,
	\begin{subequations}
		\label{even2}
		\begin{align}
		& |\widetilde{U}_0|^2 = z_0^2 ,|\widetilde{U}_{N/2}|^2=z_{N/2}^2\\
		&|\widetilde{U}_k|^2 =|\widetilde{U}_{N-k}|^2=  (z_k^2 + z_{N-k}^2)/2
		\end{align}
	\end{subequations}
	for $1\leq k \leq N/2-1$
	and $\widetilde{U}=\Lambda^Hz$, the assertion can be proved in a similar way and is omitted.
	
	When $N$ is odd, the proof is similar and one just needs to modify \dref{even} and \dref{even2} as 
	\begin{align*}
	&z_0^2 = |\widetilde{U}_0|^2,~
	z_k^2=2\times({\rm Re}(\widetilde{U}_k))^2,~
	z_{N-k}^2=2\times({\rm Im}(\widetilde{U}_k))^2
	\end{align*}
	for $1\leq k \leq (N-1)/2$ and
	\begin{align*}
	|\widetilde{U}_0|^2 = z_0^2,~
	&|\widetilde{U}_k|^2 =|\widetilde{U}_{N-k}|^2=  (z_k^2 + z_{N-k}^2)/2.
	\end{align*}
	for $1\leq k \leq (N-1)/2$.
	
	\subsection{Proof of Theorem  \ref{thm3}}
	The conclusion 1) is straightforward.
	
	Given any $|U|^2$ with its elements satisfying $|U_k|^2=|U_{N-k}|^2, 1\leq k \leq N \!-\!1$, 
	we have $\widetilde{S}(1)|U|^2 = \widetilde{S}(0)|U|^2$.
	For any given $\gamma\in\mathbb{C}$,  let $|U|^2 \in \widetilde{\mathscr{X}}^\gamma(r)$, which mean that 
	$\widetilde{S}(\gamma)|U|^2=r$, namely, $\gamma \widetilde{S}(1)|U|^2 + (1-\gamma)  \widetilde{S}(0)|U|^2=r $.
	It follows that $\widetilde{S}|U|^2=r$ due to $\widetilde{S}(1)=\widetilde{S}$. 
	This means that $\widetilde{\mathscr{X}}^\gamma(r)\subset \widetilde{\mathscr{X}}(r)$.
	Conversely, let $|U|^2 \in \widetilde{\mathscr{X}}(r)$, namely, $\widetilde{S}|U|^2 =r$.
	This derives that $\gamma \widetilde{S}(1)|U|^2 + (1-\gamma)  \widetilde{S}(0)|U|^2=r $.
	We have $\widetilde{\mathscr{X}}(r)\subset\widetilde{\mathscr{X}}^\gamma(r) $. 
	We proved that $\widetilde{\mathscr{X}}(r)=\widetilde{\mathscr{X}}^\gamma(r)$ and  accordingly		$\widetilde{\mathscr{Z}}^\gamma(r)=\widetilde{\mathscr{Z}}(r)$ and $\widetilde{\mathscr{U}}^\gamma(r)=\widetilde{\mathscr{U}}(r)$.
	This proves the conclusion 2).
	

	\subsection{Proof of Theorem  \ref{thm5}}
	
	For proving Theorem  \ref{thm5}, it suffices to show  the identity
	$$\widetilde{\Lambda}^H \widetilde{S}(1/2) \widetilde{\Lambda} = \widetilde{S}(1/2),$$
	which holds since $$
	Q_i^H 
	\begin{bmatrix}
	\cos(k\varpi)&0\\
	0&\cos(k\varpi)
	\end{bmatrix} 
	Q_i
	=\begin{bmatrix}
	\cos(k\varpi)&0\\
	0&\cos(k\varpi)
	\end{bmatrix} 
	$$
	for an arbitrary  $2\times 2$ dimensional unitary matrix $Q_i,i=1,\cdots,(N-2)/2$ with an even $N$ ($i=1,\cdots,(N-1)/2$  with an odd $N$)   and $k\geq 0$.
	
	\subsection{Proof of Theorem  \ref{thm6}}
	The proof is straightforward and is omitted.
	
	\section*{Appendix B}
	
	\renewcommand{\thesection}{B}

	\setcounter{equation}{0}
	\setcounter{thm}{0}
	\renewcommand{\thethm}{B\arabic{thm}}
	\setcounter{lem}{0}
	\renewcommand{\thelem}{B\arabic{lem}}
	\setcounter{rem}{0}
	\renewcommand{\therem}{B\arabic{rem}}
	
	This appendix contains one technical lemma.
	\begin{lem}
		\label{thma4}
		(\citet[Theorem 3.1]{Gray2006};\citet{Tee2007})
		Denote the circulant matrix $B$ generated by a row vector $b=[b_0,b_1,\cdots,b_{N-1}]$
		by
		\begin{align*}
		B={\rm circ}(b)
		\eq \left[
		\begin{array}{ccccc}
		b_0&b_1&\ddots&b_{N-2}&b_{N-1}\\
		b_{N-1}&b_0&\ddots&b_{N-3}&b_{N-2}\\
		\ddots&\ddots&\ddots&\ddots&\ddots\\
		b_2&b_3&\ddots&b_0&b_1\\
		b_1&b_2&\ddots&b_{N-1}&b_0
		\end{array}
		\right].
		\end{align*}	
		Then $B$ has unit eigenvectors
		\begin{align*}v^{(m)}&=\frac1{\sqrt{N}}\big[1,\exp(-j\varpi m),\cdots,\exp(-j\varpi (N\!-\!1) m)\big]^T, \\ m&=0,\cdots,N\!-1,\end{align*}
		where $\varpi=2\pi/N$ and $j$ is the imaginary unit ($j^2=-1$),
		and the corresponding eigenvalues
		\begin{align*}
		\tau^{(m)}&=\sum_{k=0}^{N-1}b_k\exp(-j mk\varpi)\\
		&=b[1,\exp(-j m\varpi),\cdots,\exp(-j m(N-1)\varpi)]^T
		\end{align*}
		and can be expressed by
		\begin{align*}
		B=A~\!\diag([\tau^{(0)},\cdots,\tau^{(N-1)}])A^{H}
		\end{align*}
		where $A=[v^{(0)},\cdots,v^{(N-1)}]$ is unitary and $A^H$ denotes the complex conjugate transpose of $A$. 
	\end{lem}
	

\end{document}